\documentclass[final,times,twocolumn]{elsarticle}

\usepackage{siunitx}
\usepackage{amssymb}
\usepackage{amsmath}
\usepackage{hyperref}
\usepackage{graphicx} 
\usepackage{subfigure} 
\usepackage{subcaption}
\usepackage{lineno}
\usepackage[separate-uncertainty=true]{siunitx}

\journal{Nuclear Inst. and Methods in Physics Research, A}

\begin{document}

\begin{frontmatter}



\title{Production and manipulation of stable frozen hydrogen filaments} 

\author[ikp]{Jost Froning\corref{cor1}}
\cortext[cor1]{Corresponding author}
\ead{j_fron03@uni-muenster.de}
\author[ikp]{Christian Mannweiler}
\author[ikp]{Eva-Maria Hausch}
\author[ikp]{Anna-Luna Hannen}
\author[ikp]{Alfons Khoukaz}

\affiliation[ikp]{organization={Institut für Kernphysik, Universität Münster},
            addressline={Wilhelm-Klemm-Straße 9}, 
            postcode={48149},
            city={Münster}, 
            country={Germany}}

\begin{abstract}
Frozen hydrogen filaments or droplets/pellets in vacuum are of great interest as target for many experiments at hadron and lepton accelerators as well as at high-power laser systems. Especially in case of large distances between the position of target beam generation and the interaction point with an accelerator or laser beam, a high target beam stability in space and time is crucial. 
Here we present recent results on the long-term stability of frozen hydrogen filaments in vacuum, which have been obtained using a new cryogenic target generator. 
It could be shown that the trajectory of the frozen hydrogen target beam with a diameter of \SI{10}{\micro \m} remains stable for over 60 hours, with angular fluctuations below \SI{0.08}{\degree}.
Furthermore, we present a novel strategy for the manipulation of the target beam, named cryobending. Here, the produced hydrogen filament is deflected by helium gas emerging from correction nozzles. We demonstrate the deflection of the hydrogen beam with one and two nozzles, achieving deflection angles up to \SI{15.7\pm 0.4}{\degree}.
The presented results open the door for further developments of this beam position system, which enables target beam adjustment without any mechanical movement of the target components itself.
Potential applications of these stable hydrogen target filaments are, e.g., the upcoming MAGIX (MAinz Gas Injection target eXperiment) or the planned $\mathrm{\bar{P}ANDA}$ (anti-Proton ANnihilation at DArmstadt) experiment.

\end{abstract}




\end{frontmatter}



\section{Introduction}\label{intro}
Hydrogen targets are used for a variety of experiments at hadron and lepton accelerators or at high-power laser and radioisotope beam facilities \cite{designreport} \cite{anke} \cite{cosy11} \cite{pelletpaper} \cite{magixpaper} \cite{düsseldorf} \cite{RI}. For the proposed $\mathrm{\bar{P}ANDA}$ (anti-Proton ANnihilation at DArmstadt) experiment at FAIR (Facility for Antiproton and Ion Research), both a cluster-jet and a pellet/droplet target are foreseen \cite{designreport}. The requirements for these targets are high. They have to deliver localized material in the ultra-high accelerator vacuum without the use of windows with a design goal for the areal density of \SI{4e15}{atoms/cm^2} in a distance of more than \SI{2}{\m} to the target beam production nozzle to allow for the construction of a $4\pi$ detector. Since gas jets are unable to provide such target thicknesses at such distances, the solution is to provide a jet of condensed matter particles (clusters, pellets or droplets) with sizes in the nm to \si{\micro\metre} range \cite{designreport}. A cluster-jet target forces high pressure (usually \SI{5}{\bar} to \SI{18}{\bar}) cryogenic fluid through a convergent-divergent Laval nozzle. The fluid cools down adiabatically and can condense into clusters \cite{designreport}. Different to this, a droplet target works with an aperture nozzle at pressures of about \SI{1}{\bar}. The cryogenic cylindrical liquid jet is broken up by a piezoelectric element into single spherical droplets which form a continuous stream. When the droplets freeze out due to the low vacuum pressure, they are referred to as pellets.\\
Which target is used depends on the required experimental conditions. For example, if the target thickness is desired to be adjustable and to show no time structure, a cluster-jet target is preferred. If a precise knowledge of the primary and secondary vertex is requested, a droplet or pellet target is better suited. The droplet target can also be operated without the piezo element used to break up the liquid stream into droplets, which removes the time structure in the target density. The target beam then consists of a single frozen filament instead of droplets. This alternative operation mode is applied for the results reported here. \\
Such a filament target is discussed to be an interesting and complementary alternative for some planned experiments, for example for the gas-jet target of the future MAGIX (MAinz Gas Injection target eXperiment) experiment at MESA (Mainz Energy-recovering Superconducting Accelerator). In detail, such a frozen hydrogen could allow for similar high local target thicknesses of \SI{e19}{atoms/cm^2}, but at much lower gas flow rates through the jet nozzle ($\mathrm{\SI{0.2}{l_n/min}}$ instead of $\mathrm{\SI{40}{l_n/min}}$ \cite{magixpaper}), resulting in improved vacuum conditions. \\
In order to investigate the feasibility of using cryogenic hydrogen filament target streams for both the $\mathrm{\bar{P}ANDA}$ and the MAGIX experiment, a new target device was built up (cf. chapter \ref{setup}) and used for detailed studies. 
The crucial point for the use of such a target in large-scale experiments is the stability of the target beams. 
Therefore, main subject of this manuscript is the long-term stability measurement of the hydrogen beam to demonstrate the potential of the filament target for future experimental applications. 
Furthermore, a new method to manipulate the target beam trajectory is developed.\\%
If the target beam has to travel a long distance to the interaction point as for the $\mathrm{\bar{P}ANDA}$ experiment, it must be ensured that the beam reaches the intended interaction point during the entire measuring operation and that it is possible to adjust the target beam position if necessary. 
This can be achieved by different methods. For the $\mathrm{\bar{P}ANDA}$ cluster-jet target, for example, a complex spherical alignment system is implemented which allows to tilt the central part of the cluster target generator relative to the nominal cluster-jet flight direction by a step motor system \cite{designreport}. The technical challenge here was the requirement that the centre of rotation must coincide with the narrowest cross-section of the cluster nozzle so that a tilting of the cluster source and thus a change in the nozzle angle is not accompanied by a transverse displacement of the nozzle. Although this arrangement could be designed with virtually no play, another challenge was to make this tilting device vacuum-tight, as the use of vacuum bellows was not possible for geometric reasons. 
A complementary mechanical solution to manipulate the trajectory of a droplet/pellet beam is used for the WASA (Wide Angle Shower Apparatus) pellet target device. Here, two coordinate tables are used to align the pellet stream along the path defined by an orifice (skimmer), the vacuum tubes including the interaction point in the scattering chamber, and the dump. The first table is used to direct the droplet stream through a vacuum-injection capillary and the second one moves and tilts the whole pellet generator in a way that the pellet stream passes the interaction point \cite{pelletpaper}. Whilst this design allows very precise mechanical adjustability and is compatible with ultra-high vacuum conditions thanks to the use of vacuum bellows, this arrangement is sensitive to mechanical vibrations which can be transmitted to the pellet jet.
These two methods naturally come along with additional mechanical components required increasing the size and the installation complexity of the target systems. These are, for example, motor drives and bellows that allow the entire target source to be moved/tilted in a vacuum-tight manner. 
An approach with less additional mechanical components is used for the former jet target for the Experiment 835 at the Fermilab Antiproton Accumulator \cite{gasjetFermi}. Here, instead of moving/tilting the whole target source it is possible to only tilt the target nozzle. For this purpose, the nozzle in a holder structure is connected to the cryogenic cooler stage via a copper spool so that the nozzle can be tilted \cite{gasjetFermi}. The advantage that only the nozzle is movable is accompanied by a significant loss of cooling power due to the additional copper spool.\\
To circumvent the disadvantages mentioned above, a novel method called cryobending was developed and which is the subject of this manuscript. The underlying idea is to mechanically decouple the jet beam manipulation from the actual target generator. The approach here is to deflect the target beam after leaving the target generator by transverse gas jets. Here, no adaptions at the target source itself are necessary since the redirecting devices can be implemented independently of the target source. Furthermore, there is no loss of cooling power due to an additional transfer part between the cold stage and the target nozzle.\\

\section{Setup of the cryogenic droplet target}\label{setup}
\begin{figure}[h!]
  \centering
  \includegraphics[width=0.48\textwidth]{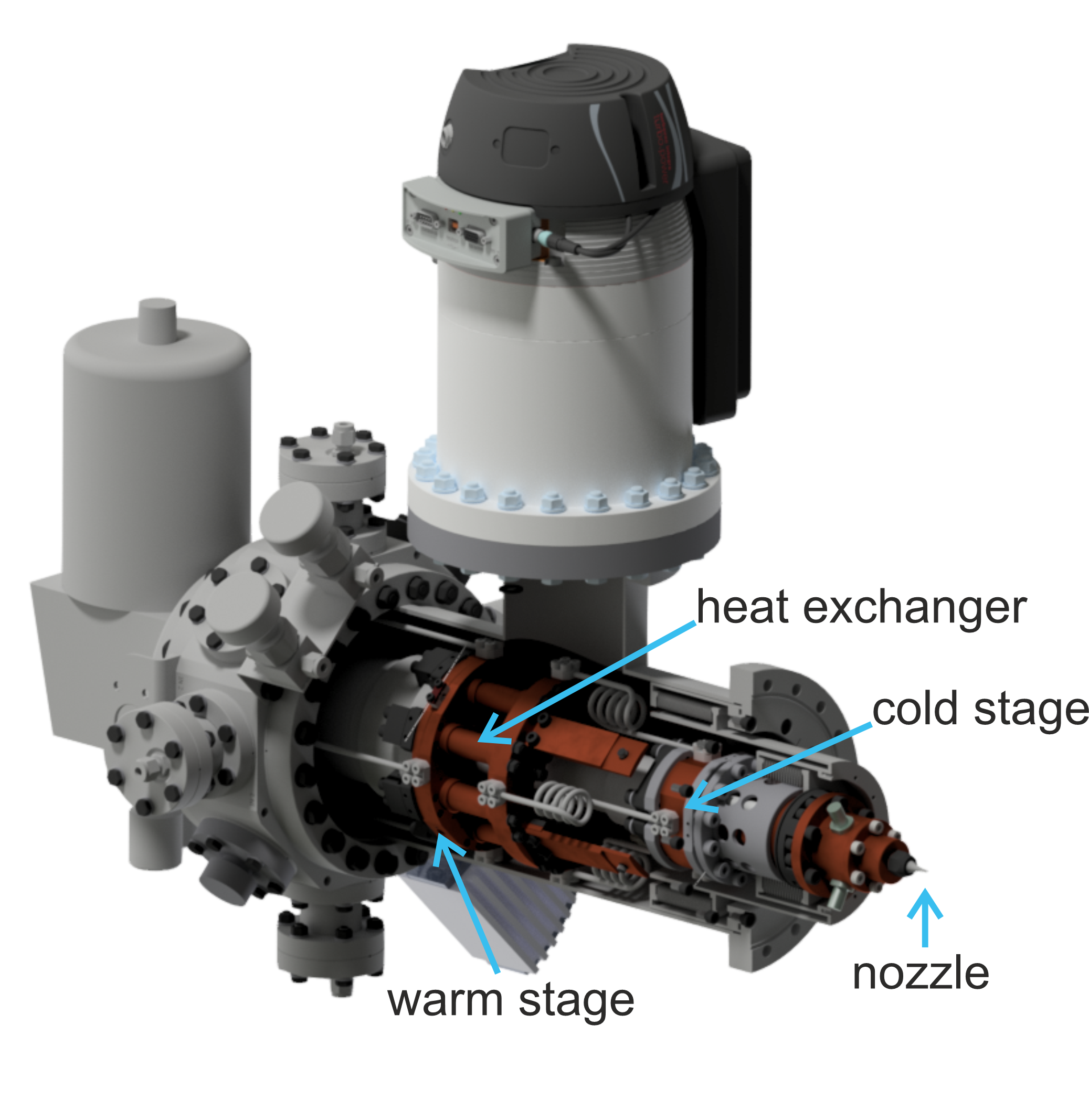}
  \caption{CAD model of the Münster droplet target: The cooling is provided by a two-stage cold head and is supplied to the process gas, i.e. hydrogen, by heat exchangers. The target has a length of roughly \SI{70}{\centi\metre} from the upper end of the cold head to the nozzle.}
  \label{Aufbau}
\end{figure}
As all the Münster targets \cite{magixpaper} \cite{magixtarget}, the Münster droplet target is based on a two-stage cold head which can reach temperatures (without gas supply) of $T_\mathrm{WS}=\SI{31}{\K}$ at the warm stage and down to $T_\mathrm{CS}=\SI{9}{\K}$ at the cold stage. The specifications of all relevant droplet target components can be found in table \ref{komponenten}. The cold head is connected via flexible high pressure lines to a compressor and works according to the Gifford-McMahon principle. For the cooling of the hydrogen target gas, a new, innovative design was developed, which is depicted in figure \ref{Aufbau}.
Instead of copper pipes welded to the cold head as in previous targets \cite{magixtarget}, the cooling of the hydrogen gas is ensured by compact heat exchangers at both the warm and the cold stage, which are copper cylinders with holes drilled lengthwise to maximize surface contact between the cold copper parts and the gas. The copper cylinders have the advantage that they can be exchanged individually, which makes them easy to clean according to the required gas purity. To minimize the loss of cooling power to the environment, the cold head is surrounded by the vacuum of the insulation chamber, which is generated by a turbo pump and a rotary vane pump and reaches pressures of $\SI{e-7}{\milli\bar}$. Additionally, the warm stage is connected to a heat shield surrounding the cold stage to minimize the impact of thermal radiation. As the cold head has no intrinsic temperature regulation and always cools down to end temperature, heating cartridges and temperature diodes are installed at both the warm and cold stage. The diodes monitor the current temperature and the cartridges emit a certain heating power so that the desired temperature set point is reached and held in a closed PID loop. \\

\begin{figure}[h!]
  \centering
  \includegraphics[width=0.5\textwidth]{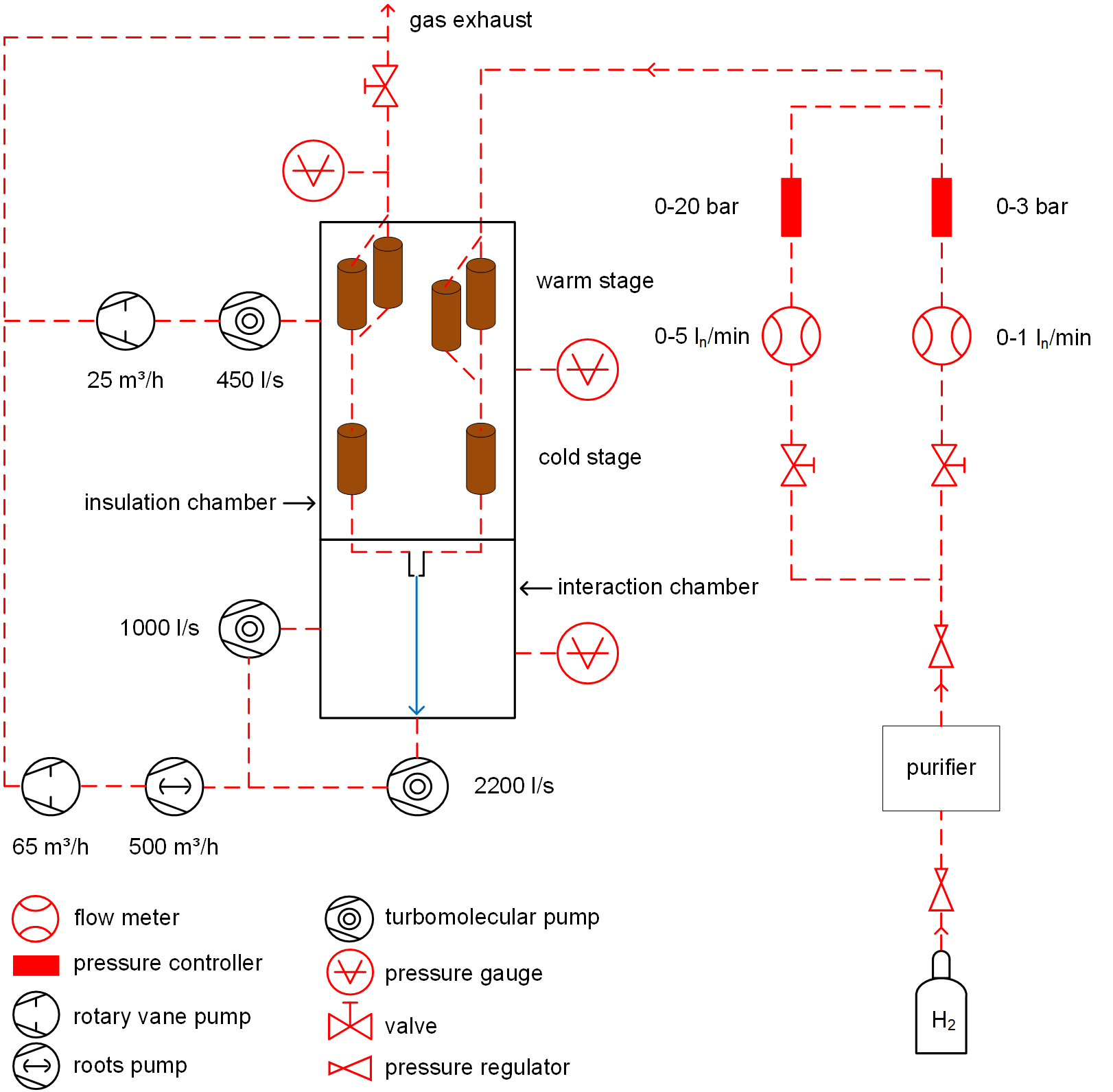}
  \caption{Schematic gas and vacuum system of the Münster droplet target: The hydrogen gas from a gas bottle is purified and enters the target via a main line and a bypass line, both of which are equipped with a flow meter and a pressure controller. In the warm and cold stage, the gas is cooled by surface contact with the heat exchangers (brown). The hydrogen leaves the target through the nozzle into the interaction chamber as a filamentary beam (blue). The vacua in the interaction and insulation chambers are generated by the pumps at the specified pumping speeds.}
  \label{Gassystem}
\end{figure}
A schematic sketch of the target gas system is shown in figure \ref{Gassystem}. The hydrogen gas for the target system is supplied by a $\SI{200}{\bar}$ gas bottle with $99.999\,\%$ gas purity. 
In a first step, a pressure regulator is used to reduce the process gas pressure to a value of \SI{20}{\bar}. The purity of the gas is further increased to ppb level by a palladium-hydrogen purifier. A second pressure regulator following the purifier sets the gas pressure to a value of roughly \SI{5}{\bar}. The gas can be then fed into the target via a main line and a bypass line, both equipped with a flow meter and a pressure controller. The bypass line allows for a higher flow rate required to liquefy the hydrogen. Inside the target, the gas is pre-cooled in the warm stage by two heat exchangers. Via a helical pipe, the gas is guided to the cold stage where it is further cooled down by one heat exchanger to liquid hydrogen conditions. From there, the hydrogen is fed to a piezo head (not in operation here) and the target nozzle which consists of a copper body and a welded-on platinum-iridium orifice with an aperture diameter of \SI{10}{\micro\metre}. The hydrogen enters the vacuum of the interaction chamber after passing through the nozzle and freezes out due to the low vacuum chamber pressure, so that a filamentary beam with a diameter of roughly \SI{10}{\micro\metre} can be observed. The vacuum of the interaction chamber is generated by two turbo pumps in parallel and subsequently by a roots pump and a rotary vane pump. Vacuum pressures in the order of $\SI{e-4}{\milli\bar}$ are achieved during the operation at liquid hydrogen conditions and are mainly due to the fact that the generated hydrogen filament is completely dumped in this chamber. An outlet line (which also runs through heat exchangers) enables the release of hydrogen in the case of overpressure due to setting the target from liquid to gaseous conditions. A baratron on the outlet line allows the gas line pressure to be measured.\\ 

\begin{table}[t]
  \centering
  \caption{Specification of the components for cooling, controlling and pumping the Münster droplet target.}
  \scalebox{0.8}{
  \begin{tabular}{c|c}
    Component & Specification \\ \hline
    Cold head & Leybold Coolpower 10 MD \\
    Compressor & Leybold Coolpak 6000 H \\
    Temperature control & Lakeshore 336 \\
    Temperature diodes & Lakeshore DT-670 \\
    Heating cartridges & Lakeshore HTR-25-100 \& HTR-50 \\
    Hydrogen gas purifier & MegaTorr PS7-PD \\
    Pressure controller & Brooks Instruments SLA 5810 \\
    Flow meter & Brooks Instruments 5850E \\
    Gas pressure gauge & MKS Baratron 722B \\
    Vacuum gauge & Leybold IONIVAC ITR90 \\
    & Leybold CERAVAC CTR100 \\
    Rotary vane pump & Leybold TRIVAC D 65 B \& D 25 B \\
    Roots pump & Leybold RUVAC WSU 501 \\
    Turbomolecular pump & Leybold TURBOVAC 450i \& 1000C \\
    & Pfeiffer TPH 2200 \\
  \end{tabular}}
  \label{komponenten}
\end{table}

\begin{figure}[h!]
  \centering
  \includegraphics[width=0.5\textwidth]{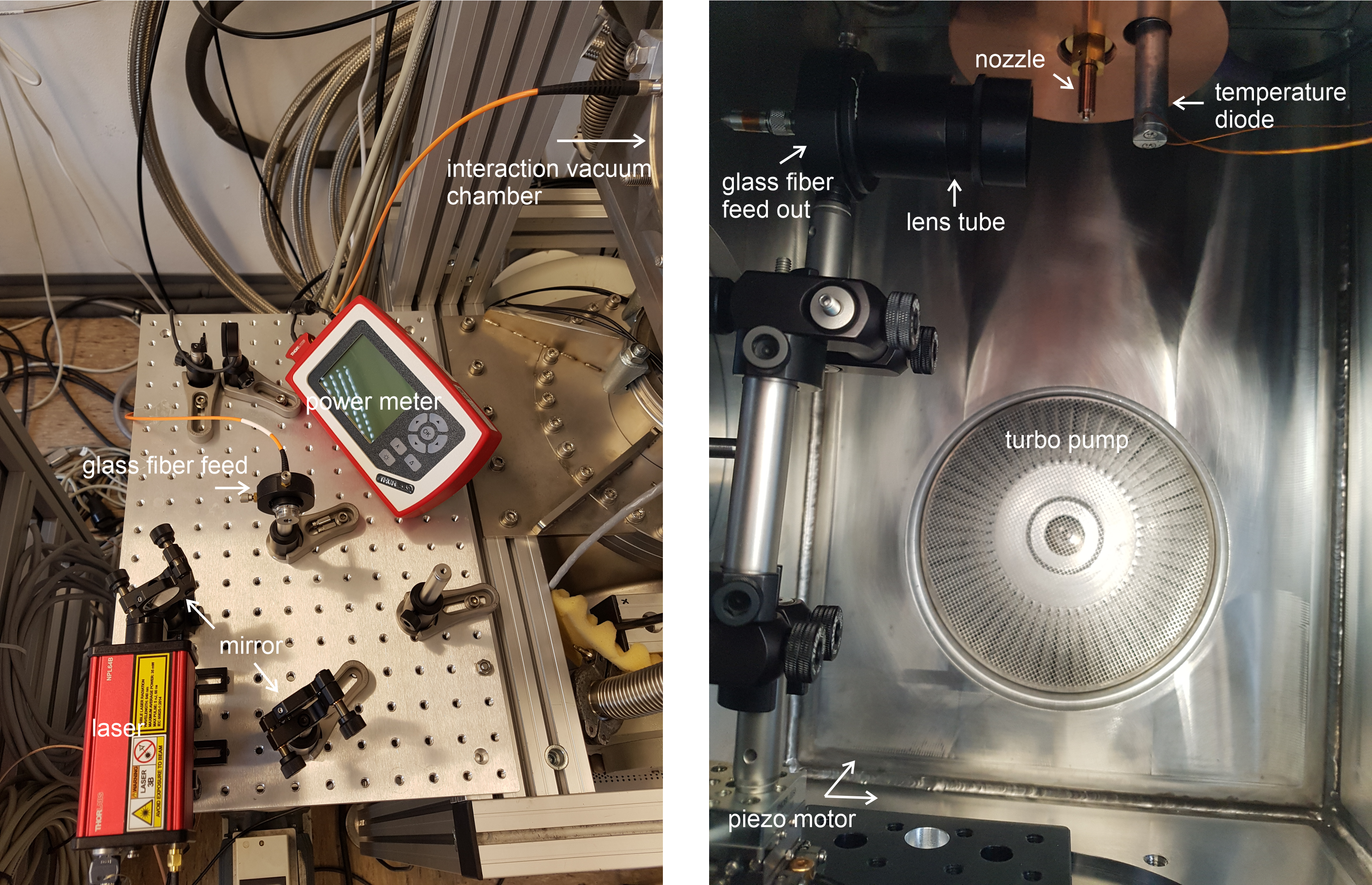}
  \caption{Optical monitor system of the Münster droplet target. The distance between the nozzle exit and the lower wall of the vacuum chamber amounts to roughly \SI{25}{\cm}.}
  \label{monitor}
\end{figure}
To analyse the stability of the hydrogen target beam, the monitoring system in figure \ref{monitor} is applied. The pulses of a ns laser (repetition rate \SI{1}{\MHz} to \SI{10}{\MHz}, pulse witdh \SI{5\pm 1}{\ns} to \SI{39\pm 3}{\ns}) are fed via two mirrors into an optical fiber which is connected to the interaction chamber (left picture). There, the pulses are coupled into a UHV feedthrough and guided in a vacuum onto a lens tube. 
The lens focuses the light pulses on the hydrogen beam emerging from the target nozzle and is adjusted to the beam via a piezo motor device. At the location of the target stream, the laser has a diameter of less than \SI{1}{\milli\metre}. The expected target beam fluctuation is much smaller than the laser beam diameter, so that continuous readjustment of the motor is not necessary.
The illuminated target beam is observed by a telescope objective and a camera (exposure time down to \SI{12}{\micro\s}) mounted outside of the interaction chamber. At a distance of \SI{2.8}{\centi\metre} between laser focus and target nozzle, the camera field of view is \SI{750}{\micro\metre} x \SI{560}{\micro\metre} with a resolution of \SI{0.58}{\micro\metre /px}. The optical monitor system is calibrated by placing a wire with a known diameter in the beam path at the expected position of the hydrogen beam. \\

The hydrogen beam might leave the nozzle not perfectly perpendicular to the ground due to imperfections of the aperture shape or small impurities, preventing the use of the target at large distances of several meters, for example. For previous droplet/pellet targets, the solution was to align the whole system mechanically with two coordinate tables \cite{pelletpaper} (cf. chapter \ref{intro}).
We propose a novel strategy, named cryobending, where blowing gas on the hydrogen beam redirects it to the desired direction. \\
\begin{figure}[h!]
  \centering
  \includegraphics[width=0.45\textwidth]{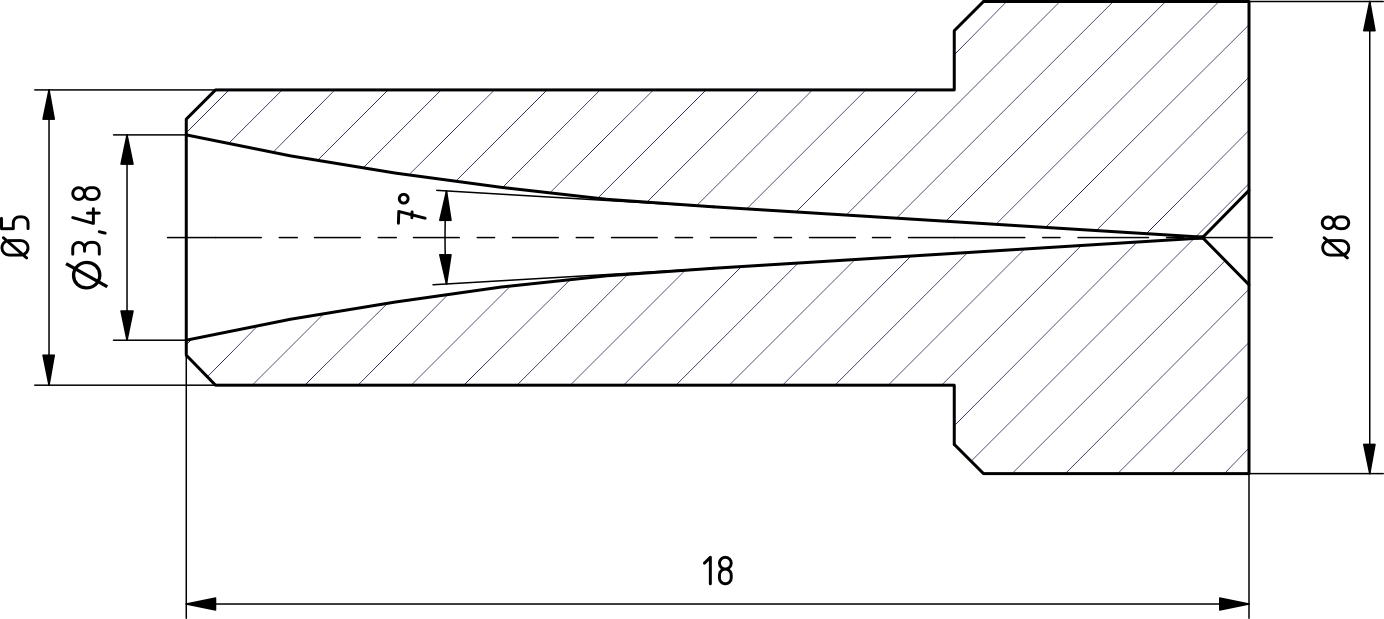}
  \caption{Technical drawing of the used Laval nozzles with a convergent inlet (right) and a divergent outlet (left). All length and diameter information are given in units of mm.}
  \label{düse}
\end{figure}
To demonstrate the feasibility of this technique, we implemented correction nozzles in the interaction chamber of the droplet target beneath the hydrogen nozzle. Laval-type nozzles are used for cryobending (see chapter \ref{2}), among others. The geometry of these nozzles is shown in figure \ref{düse}. It consists of a convergent $\SI{90}{\degree}$ inlet cone and a divergent outlet cone which is divided into two zones, the first one with an opening angle of $\SI{7}{\degree}$ and a length of $\SI{8}{\mm}$, followed by the second one with a spherical shape with a radius of $\SI{60}{\mm}$ and a length of $\SI{9}{\mm}$. The narrowest inner diameter of the Laval nozzles used here is \SI{90}{\micro\metre}.\\

\section{Stability of the cryogenic target beam}
The stability of the cryogenic target beam is analysed repeatedly by taking 100 images with 4 frames per second of the hydrogen jet.
Thus, the beam fluctuations are analysed in short intervals of 25 s each.
Its position is determined by fitting each intensity profile and averaging the resulting positions. As a measure of stability, the standard deviation of the positions is specified. \\
To investigate the long term stability of our target, we performed a 100-hour measurement in which the target was operated at liquid hydrogen conditions of \SI{16}{\kelvin}, measured directly at the nozzle holder, which is coupled to the cold stage of the cryogenic head, and \SI{1.2}{\bar} without piezo excitation in order to produce a continuous frozen filament. 
\begin{figure}[h]
  \centering
  \includegraphics[width=0.5\textwidth]{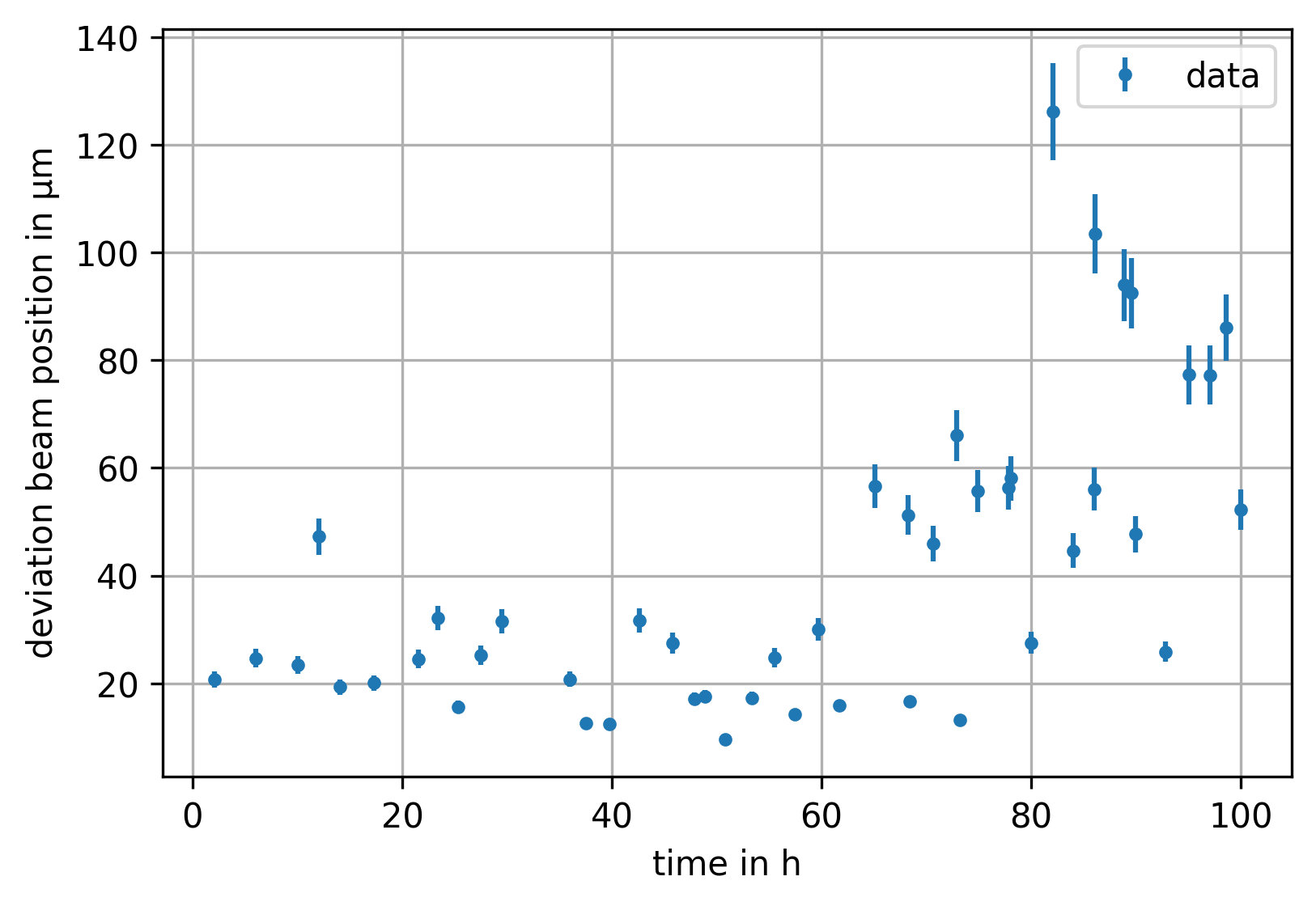}
  \caption{Standard deviation of the hydrogen beam position during the 100-hour measurement.}
  \label{100h}
\end{figure}
During this 100-hour operation, image series of the beam as explained above were captured. The resulting standard deviations $\sigma$ of the beam positions are plotted over time in figure \ref{100h}. The shown uncertainty of the standard deviation is given as $\left(2\left(n-1\right)\right)^{-1/2}\cdot \sigma$ with $n=100$. The first 60 hours of the measurement reveal a very stable behavior with positions deviations below \SI{40}{\micro\metre} at a distance of \SI{2.8}{\cm} behind the nozzle, corresponding to angular fluctuations below \SI{0.08}{\degree}. Nevertheless, the fluctuation of the beam position increases after roughly 60 hours. The measurement was ended after 100 hours, so that there is no statement possible whether the fluctuation will continue to increase or whether it will stabilize again.\\
After heating up to \SI{293}{\kelvin} and cooling down again without venting the target, the same stabilities are observed as at the beginning of the long term measurement. It is assumed that external gases such as nitrogen or oxygen freeze on the target nozzle during the measurement and block it slightly, impairing stability. By heating the target, this blockage evaporates, restoring the beam stability.
The deviations during the first 60 hours are interpreted as intrinsic target performance and the deviations after 60 hours are due to external gases entering the target chamber, since for this feasibility study an O-ring sealed chamber is used. For the application in accelerator experiments, the target chamber should be equipped with UHV-seals.\\
The beam deviations observed here range from \SI{10}{\micro\metre} to \SI{126}{\micro\metre} during the long term measurement with the images captured at a distance of \SI{2.8}{\centi\metre} to the nozzle exit. This implies via the intercept theorem deviations of \SI{0.73}{\milli\metre} to \SI{9.5}{\milli\metre} in a distance of \SI{2.1}{\metre} to the nozzle. This will be, for example, the distance of the target nozzle to the interaction point of the $\mathrm{\bar{P}ANDA}$ experiment at FAIR \cite{grieser}, a possible future application example of our droplet target at a storage ring experiment, where the high stability at large distances is mandatory. Therefore, to ensure that the target beam reaches the desired position, a beam positioning system (cryobending) should be applied.\\
Furthermore, preparations are currently ongoing to extend the target beam line so that measurements of the filament and its stability at larger distances from the nozzle become possible. Additionally, we already performed stability measurements at a distance of \SI{17.5}{\cm} from the filament nozzle. Here, transversal fluctuations between \SI{33}{\micro\metre} and \SI{212}{\micro\metre} are observed, which would correspond to lateral deviations between \SI{0.4}{\milli\metre} and \SI{2.5}{\milli\metre} at a distance of \SI{2.1}{\metre} from the nozzle. The frozen hydrogen target beam thus appears to stabilize itself due to gravitational forces pulling it downwards. Furthermore, a break-up of the frozen filament target beam is not observed after a distance of \SI{17.5}{\cm}. From these observations, it is plausible to assume that even if a subsequent break-up occurs there will be no fragments with random directions since there is no source of additional transversal distortion. \\
The vacuum tube at the interaction point of the future $\mathrm{\bar{P}ANDA}$ experiment is planned to have a diameter of \SI{2}{\cm} and the antiproton beam will have a minimum diameter of \SI{100}{\micro\metre} \cite{designreport}. However, by using quadrupole magnets of the antiproton accelerator HESR (High Energy Storage Ring), the (horizontal) size of the antiproton beam can be continuously adjusted to the transversal spread of the target beam. Therefore, the observed stability of the target beam after \SI{17.5}{\cm}, which corresponds to a spread in the (sub-) millimeter range at the expected $\mathrm{\bar{P}ANDA}$ interaction point, will be more than sufficient for the discussed application.\\ 
In comparison to our target system, for example, the WASA pellet target was reported to be operated for a period of four days and its pellet stream had a lateral size of \SI{2}{\milli\metre} FWHM at a distance of \SI{1.37}{\metre} downstream of the nozzle \cite{pelletpaper}. The calculated best stability of our hydrogen beam at \SI{1.37}{\metre} corresponds to a FWHM of \SI{1.1}{\milli\metre}, which is almost an improvement by a factor of two. In addition, it must be mentioned that a skimmer is used in the WASA target system to reduce the pellet rate by more than $50\,\%$, so that the beam divergence is artificially improved by cutting out pellets with a too large angle divergence. Contrary, in the measurements presented here the full target beam is used without any beam skimmers.
A period of four days of stable operation is not yet reached by our target, but longer operation times are possible if short regeneration intervals of a few hours are used, e.g. every 60 hours, to heat up and cool down the target nozzle.\\
Although both the future $\mathrm{\bar{P}ANDA}$ and the MAGIX experiment are considered to take data for periods of weeks, the presented filament target might be an interesting target for both installations. In case of $\mathrm{\bar{P}ANDA}$ there will be cycles of, e.g., \SI{1800}{\s} of continuous data taking \cite{designreport}. After this time period, a new antiproton beam will be prepared. In addition, the use of a cryopump is considered for $\mathrm{\bar{P}ANDA}$ with a possible regeneration interval of a few days. Thus, these regeneration or beam preparation times can be used for a target regeneration. Similarly, in case of MAGIX there will be measurements at different magnet spectrometer angles. Each angle setting will be typically used for a few days, so the time needed for changing the angle setting (hours) can also be used for a target regeneration. From this point of view, the operation time achieved here is already suitable for both experiments.\\

\section{Deflection of frozen hydrogen filaments}
In this section we show the theoretical description of our proposed cryobending method and first approaches to deflect the hydrogen filament with one convergent nozzle and with two Laval nozzles as a proof of principle. \\
\subsection{Theoretical description of cryobending}\label{theory}
\begin{figure}[h]
  \centering
  \includegraphics[width=0.5\textwidth]{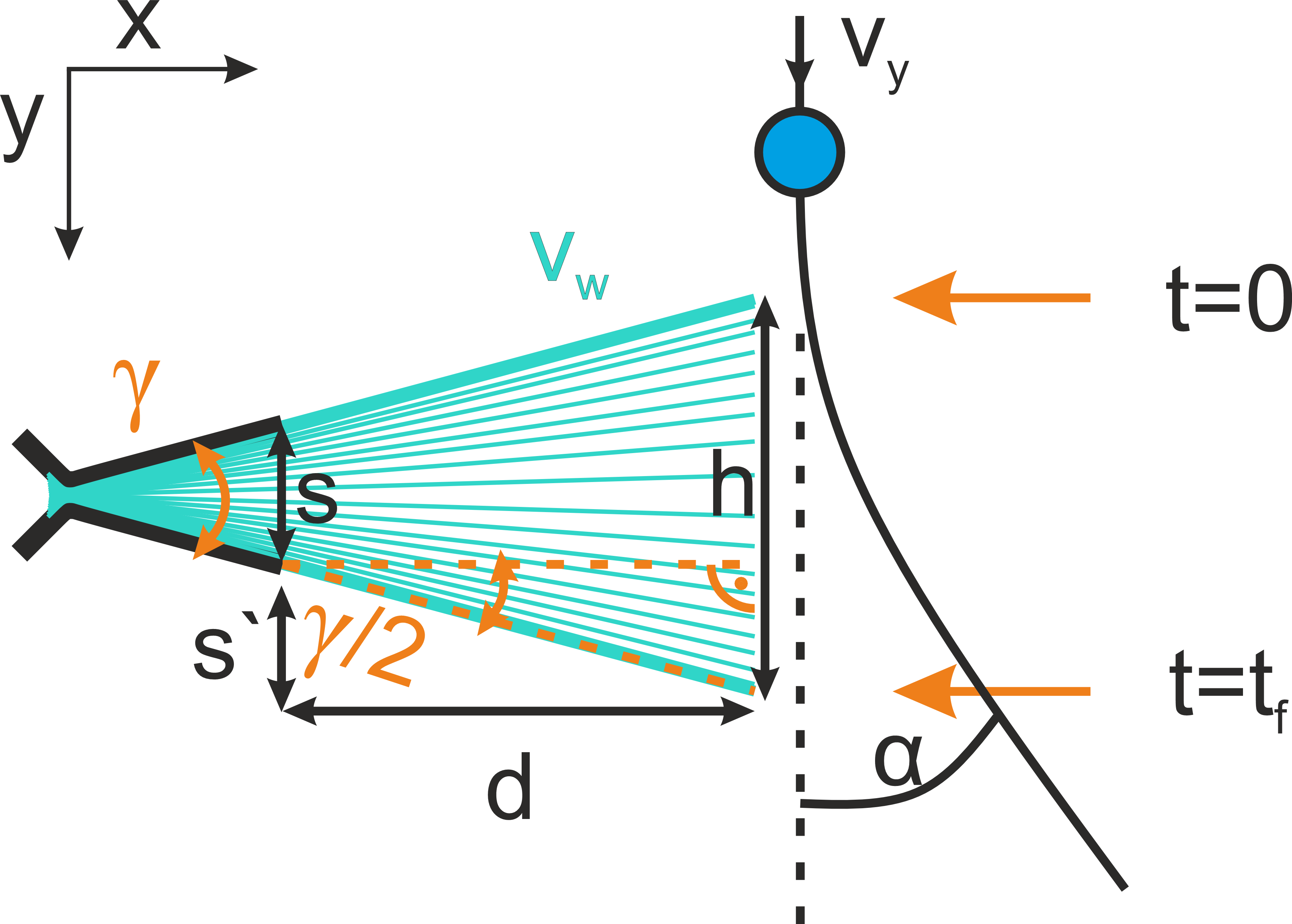}
  \caption{Schematic sketch of cryobending to derive a formula for the deflection angle $\alpha$ (not to scale).}
  \label{skizze}
\end{figure}

Cryobending describes the proposed deflection of our frozen hydrogen filament target beam by transversally blowing gas onto it. Helium is used as the deflection gas, as it does not freeze at the target nozzle at the operating temperature of \SI{16}{\kelvin}. A schematic sketch of the functional principle is shown in figure \ref{skizze}. The bending angle $\alpha$ of the redirected hydrogen beam can be calculated with the following assumptions. For simplicity, instead of a filament we assume a spherical droplet with a diameter of \SI{10}{\micro\metre} leaving the target nozzle with the velocity $v_y$ in y-direction. This velocity can be calculated via 
\begin{equation}\label{vy}
  v_y=\sqrt{\frac{2p_H}{\rho_H}}
\end{equation}
with the hydrogen pressure at the target nozzle $p_H$ and the hydrogen density $\rho_H$ \cite{varentsov}. At liquid hydrogen conditions of \SI{16}{\kelvin} and \SI{1}{\bar}, this results in typical velocities of $v_y\approx \SI{52}{\m / \s}$.
The droplet gains a velocity component $v_x(t)$ in x-direction starting at the time $t=0$ due to the interaction with the deflection gas with the speed $v_w$ from the bending Laval nozzle. At $t=t_f$, the droplet leaves the influence zone of height $h$ of the bending nozzle. The bending gas is assumed to exert a drag force \cite{drag}
\begin{equation}\label{drag}
  F_d=\frac{1}{2}\rho v_w^2 C_d A
\end{equation} 
on the droplet with the bending gas density $\rho$, the drag coefficient $C_d=0.5$ for spherical objects \cite{hoerner} and the cross-sectional area of the droplet $A=\pi r^2$. Using \eqref{drag}, the equation of motion can be written as $F_d=m\dot{v}_x$ with the droplet mass $m=4\rho_H\pi r^3/3$ and the droplet radius $r$ and is solved with the boundary conditions $v_x(0)=0$ and $v_x(t_f)=v_x$ as
\begin{equation}\label{vx}
  v_x=\frac{\rho v_w^2 C_d A t_f}{2m} \stackrel{t_f=h/v_y}{=} \frac{\rho v_w^2 C_d A h}{2m v_y}.
\end{equation}
Hence, for the bending angle
\begin{equation}\label{alpha1}
  \tan{\alpha}=\frac{v_x}{v_y} = \frac{\rho v_w^2 C_d \pi r^2 h}{2 v_y^2 \rho_H 4/3 \pi r^3} = \frac{3 \rho v_w^2 C_d h}{8 v_y^2 \rho_H r}.
\end{equation}
applies. The quantities $\rho$, $v_w$ and $h$ in this equation have to be determined. Approximating the spacial expansion of the gas jet beam to be linear, the zone height $h$ can be estimated via geometric considerations shown in figure \ref{skizze} as 
\begin{equation}\label{h}
  h=s+2s'=s+2d\tan\left(\gamma/2\right).
\end{equation}
Here, $s$ is the inner diameter of the Laval nozzle at its exit, $\gamma$ is the opening angle and $d$ is the distance of the droplet to the nozzle exit.\\
For the speed of a perfect bending gas
\begin{equation}\label{vw}
  v_w=\sqrt{\frac{2\kappa}{\kappa -1}\frac{RT}{M}}
\end{equation}
applies \cite{taschner} with the universal gas constant $R$, the inlet gas temperature $T$, the molar mass $M$ and the heat capacity ratio $\kappa$.
The bending gas density is given by dividing the mass flow rate $\dot{m}$ by the gas speed $v_w$ and the cross-section of the bending beam at the interaction area $\pi h^2/4$:
\begin{equation}\label{rho}
  \rho=\frac{\dot{m}}{v_w \pi h^2/4} = \frac{M p_N}{R T_N}\frac{q_V}{v_w \pi h^2/4}.
\end{equation}
The mass flow rate is here expressed in terms of the volume flow rate $q_V$ as well as the pressure $p_N$ and temperature $T_N$ at standard conditions \cite{grieser}. The volume flow rate is given as \cite{grieser}
\begin{equation}\label{qv}
  q_V=\pi \frac{d_N^2}{4} \frac{p}{\sqrt{M T}}\frac{T_N}{p_N}\left(\frac{2}{\kappa+1}\right)^{\frac{\kappa+1}{2\left(\kappa-1\right)}}\sqrt{\kappa R}
\end{equation}
with the inner nozzle diameter $d_N$ and the bending gas pressure $p$. With \eqref{qv}, the density is obtained as
\begin{equation}\label{rho2}
  \rho= d_N^2 p \sqrt{\frac{M}{T}} \left(\frac{2}{\kappa+1}\right)^{\frac{\kappa+1}{2\left(\kappa-1\right)}} \sqrt{\frac{\kappa}{R}} \frac{1}{v_w h^2}.
\end{equation}
Finally, \eqref{vy}, \eqref{h}, \eqref{vw} and \eqref{rho2} can be inserted in \eqref{alpha1} so that
\begin{equation}\label{alpha2}
  \tan{\alpha}=\frac{3 C_d d_N^2 \kappa }{16 r} \left(\frac{2}{\kappa+1}\right)^{\frac{\kappa+1}{2\left(\kappa-1\right)}} \sqrt{\frac{2}{\kappa-1}} \frac{1}{s+2d\tan\left(\gamma/2\right)} \frac{p}{p_H}
\end{equation}
is obtained. Without mechanical manipulation of the used target components, the bending angle can therefore be controlled primarily via the bending gas pressure $p$. Furthermore, the angle can be reduced by increasing the hydrogen pressure at the nozzle $p_H$, the droplet radius $r$ or the distance to the bending nozzle $d$. 
Plugging in typical parameters to \eqref{alpha2} as $d_N=\SI{100}{\micro\metre}$, $\kappa=5/3$ for helium at room temperature, $r=\SI{5}{\micro\metre}$, $s=\SI{3.48}{\milli\metre}$, $d=\SI{2}{\centi\metre}$, $\gamma=\SI{7}{\degree}$ and $p_H=\SI{1}{\bar}$, bending angles of $\SI{2.9}{\degree}$ to \SI{19.7}{\degree} are calculated for bending pressures of \SI{1}{\bar} to \SI{7}{\bar}.
This rough estimate shows that considerable deflections of the hydrogen beam can be achieved with our cryobending system. It should be noted that this derivation is based on simplified assumptions. The bending gas is assumed to have a constant velocity $v_w$ and a divergence which follows the inner shape of the bending nozzle. In fact, the expansion of a gas jet from a Laval nozzle shows complex ejection patterns \cite{magixpaper} so that numerical simulations have to be applied for exact knowledge of the jet beam structure in vacuum. Moreover, a detailed description of the bending of frozen filaments requires also the consideration of their solid state properties and elasticity. Nevertheless, these assumptions are valuable for a qualitative understanding of the working principle of the cryobending system presented here.\\


\subsection{First approach to cryobending}\label{1}
\begin{figure}[h!]
  \centering
  \includegraphics[width=0.5\textwidth]{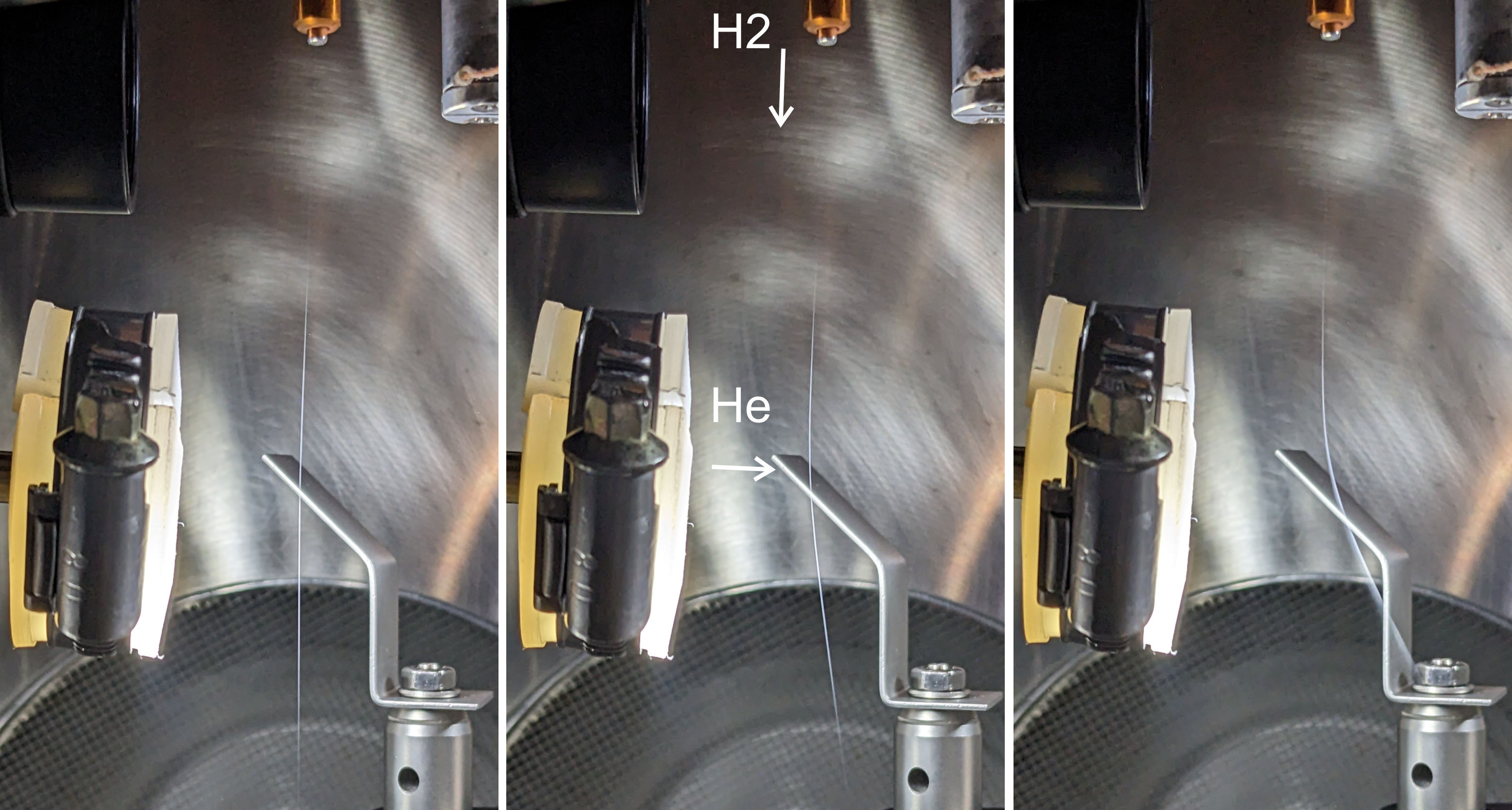}
  \caption{Deflection of the frozen hydrogen target beam by helium gas emerging from the left nozzle. The helium gas pressure is set to $\SI{1}{\bar}$, $\SI{4}{\bar}$ and $\SI{7}{\bar}$ from left to right. The vertical scale of the picture amounts to $\sim\SI{120}{\milli\metre}$.}
  \label{ersteDüse}
\end{figure}

As a first attempt of cryobending, one deflection nozzle with a convergent inlet and an inner diameter of $d_N=\SI{100}{\micro\metre}$ is implemented in the interaction chamber of the droplet target as shown in figure \ref{ersteDüse}. This bending nozzle is placed beneath the copper-colored hydrogen nozzle of the target in a vertical distance of roughly \SI{5.5}{\cm} and a horizontal distance of \SI{1.5}{\cm}. 
The bending nozzle is connected to a helium gas supply with a gas purity of $99.999\,\%$ without further purification. While this gas purity proved to be completely sufficient for the measurements presented here, the use of commercially available adsorber cartridges, which can reduce gas impurities to a level below 0.1\,ppm, could be considered for long-term measurements of e.g. weeks. The deflection nozzle is adjusted so that there is a maximum overlap between the helium jet and the target beam.

The droplet target is set to liquid hydrogen conditions of $\SI{16}{\kelvin}$ and $\SI{1}{\bar}$ so that frozen hydrogen filaments are observed in the chamber at a vacuum pressure of $p_\mathrm{ch}=\SI{9.2e-4}{\milli\bar}$. The helium gas input pressure at the deflection nozzle is set via a pressure reducer to $\SI{1}{\bar}$, $\SI{4}{\bar}$ and $\SI{7}{\bar}$ from left to right in figure \ref{ersteDüse}. As a result the chamber vacuum increases to $\SI{6.8e-3}{\milli\bar}$, $\SI{2.0e-2}{\milli\bar}$ and $\SI{5.1e-2}{\milli\bar}$ respectively.
The measurement reveals that the deflection of the cryogenic hydrogen filament by room temperature helium works well. As expected, the deflection angle increases with higher helium pressures. The hydrogen beam remains well-defined at $\SI{1}{\bar}$ and $\SI{4}{\bar}$, while a widening is observed at $\SI{7}{\bar}$. However, it appears that such high bending pressures are practically of less relevance since the resulting deflection at $\SI{1}{\bar}$ already seems to straighten the hydrogen jet so that it can move over long distances. In addition, lower bending pressures are preferable as the vacuum pressure is not affected as much as at higher pressures. Thus, in practice one can use small helium pressures to correct the path of the hydrogen filaments. Furthermore, much lower deflection gas pressures can be used if the deflection nozzle is placed closer to the target beam than shown here in this proof of principle. 
For the measurements presented here, deflection gas jets produced at room temperature are used. It is found that the quality of the deflected target beam, e.g. the filament diameter, does not appear to be affected after deflection at lower deflection gas pressures (figure \ref{ersteDüse} left and middle). This shows that complex cooling of the deflection gas is not necessary, which makes the setup simple and less susceptible to faults.
With this first achievement, further studies on cryobending can be performed.\\

\subsection{Deflection with Laval nozzles}\label{2}
\begin{figure}[h!]
  \centering
  \includegraphics[width=0.4\textwidth]{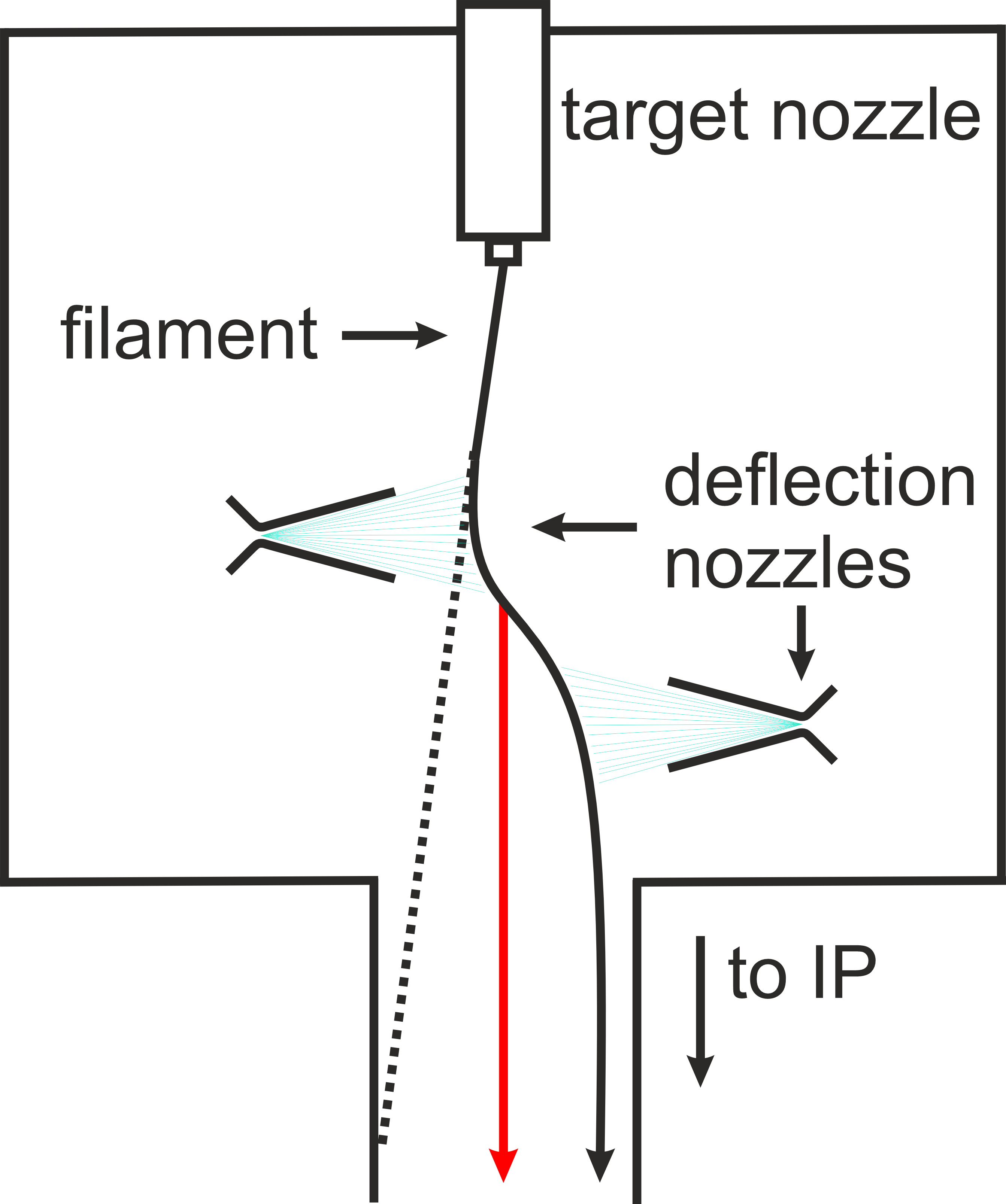}
  \caption{By using two Laval nozzles per coordinate axis perpendicular to the spread direction of the target beam, the inclined hydrogen filament can be straightened and the position controlled so that the interaction point (IP) can be reached by the target stream at any desired angle or position. By applying different bending pressures, the beam can be shifted to the center, for example (red line).}
  \label{doppelbiegung}
\end{figure}
As seen in the previous chapter, the deflection of cryogenic hydrogen works with one convergent nozzle. The next step is to use two opposing and vertically offset Laval nozzles to investigate whether full control over the beam position and trajectory can be gained as sketched in figure \ref{doppelbiegung}. In the sketch, the jet emerging from the upper target nozzle is at an angle to the nozzle axis. Without the use of bending nozzles (dashed line), the jet would hit the vacuum pipes and would not be able to reach the interaction point. By using the bending nozzles, the beam can be redirected and therefore reach the interaction point. The position can be corrected to the centre of the vacuum pipe, for example, by reducing the pressure of the left nozzle and simultaneously increasing the pressure of the right nozzle so that the beam is shifted to the left (red line). Moreover, the target beam position at the IP can be steered in order to scan, for example, an accelerator or laser beam in a later experiment.\\
In order to demonstrate the double deflection, two Laval nozzles with an inner diameter of $d_N=\SI{90}{\micro\metre}$ and an inner geometry as shown in figure \ref{düse} are implemented so that the measured and the expected deflection angle equation \eqref{alpha2} can be compared. 
\begin{figure}[h!]
  \centering
  \includegraphics[width=0.5\textwidth]{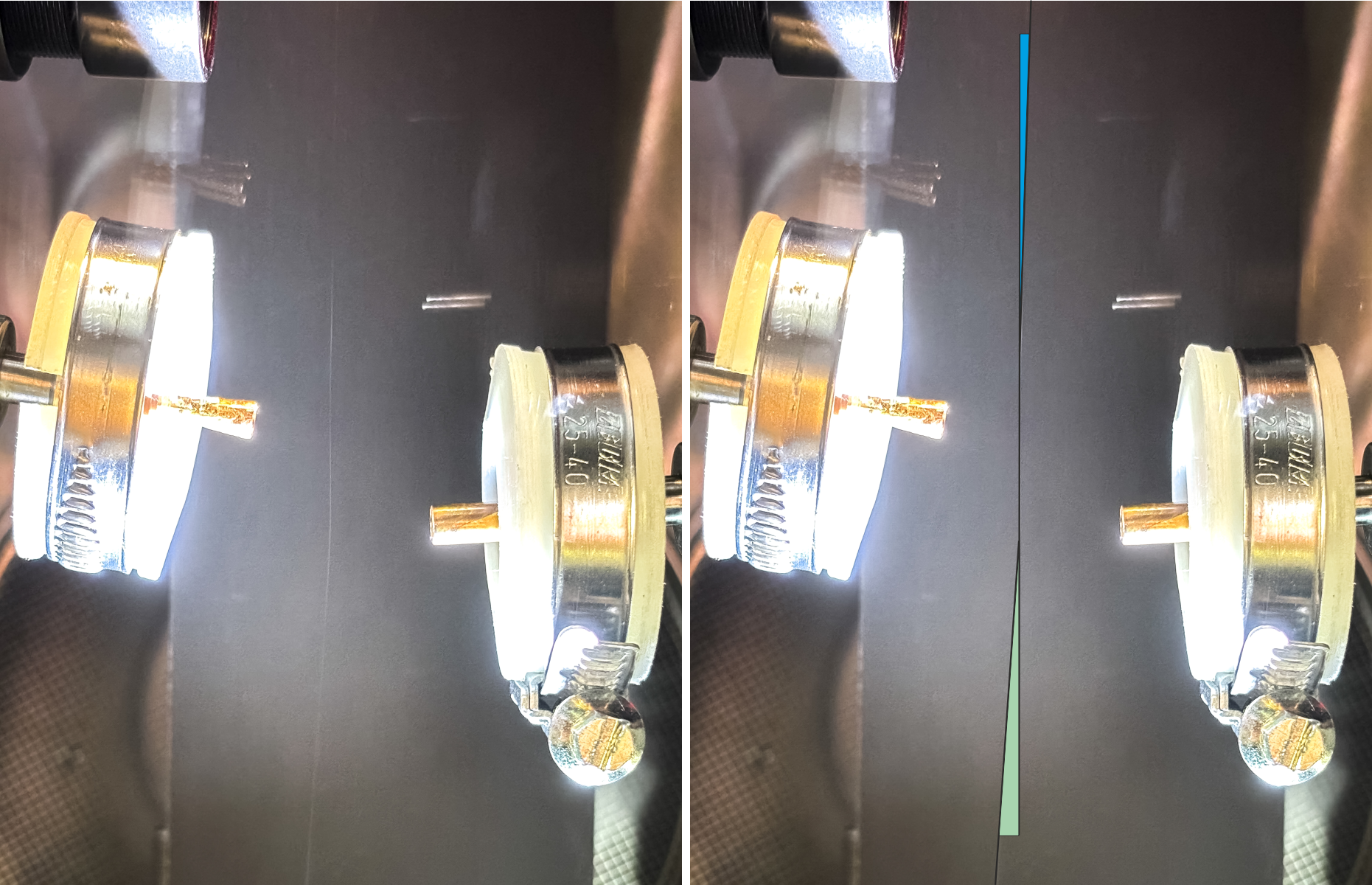}
  \caption{Double deflection of the hydrogen target beam by helium gas emerging from two Laval nozzles with a gas pressure of $\SI{1.5}{\bar}$ each. The deflection angle caused by the first nozzle is marked in blue and the angle caused by the second nozzle in green.}
  \label{zweiDüsen}
\end{figure}
Figure \ref{zweiDüsen} shows the two Laval nozzles inside the interaction chamber. Both nozzles are operated with a helium pressure of $\SI{1.5}{\bar}$. In contrast to the first test presented in section \ref{1}, a pressure controller is used here for precise knowledge of the applied pressure. The target is operated at liquid hydrogen conditions of $\SI{16}{\kelvin}$ and $\SI{1.2}{\bar}$ resulting in frozen hydrogen filaments with a diameter of $\sim \SI{10}{\micro\m}$ and a speed of $\sim \SI{56}{\m / \s}$. The first nozzle causes the skewed hydrogen filament to straighten, while the filament is tilted again by the second nozzle, with the beam maintaining its well-defined shape. Hence, the double deflection works as well as the deflection with a single nozzle. For quantitative statements, the bending angles are analysed.
\begin{figure}[h!]
  \centering
  \includegraphics[width=0.5\textwidth]{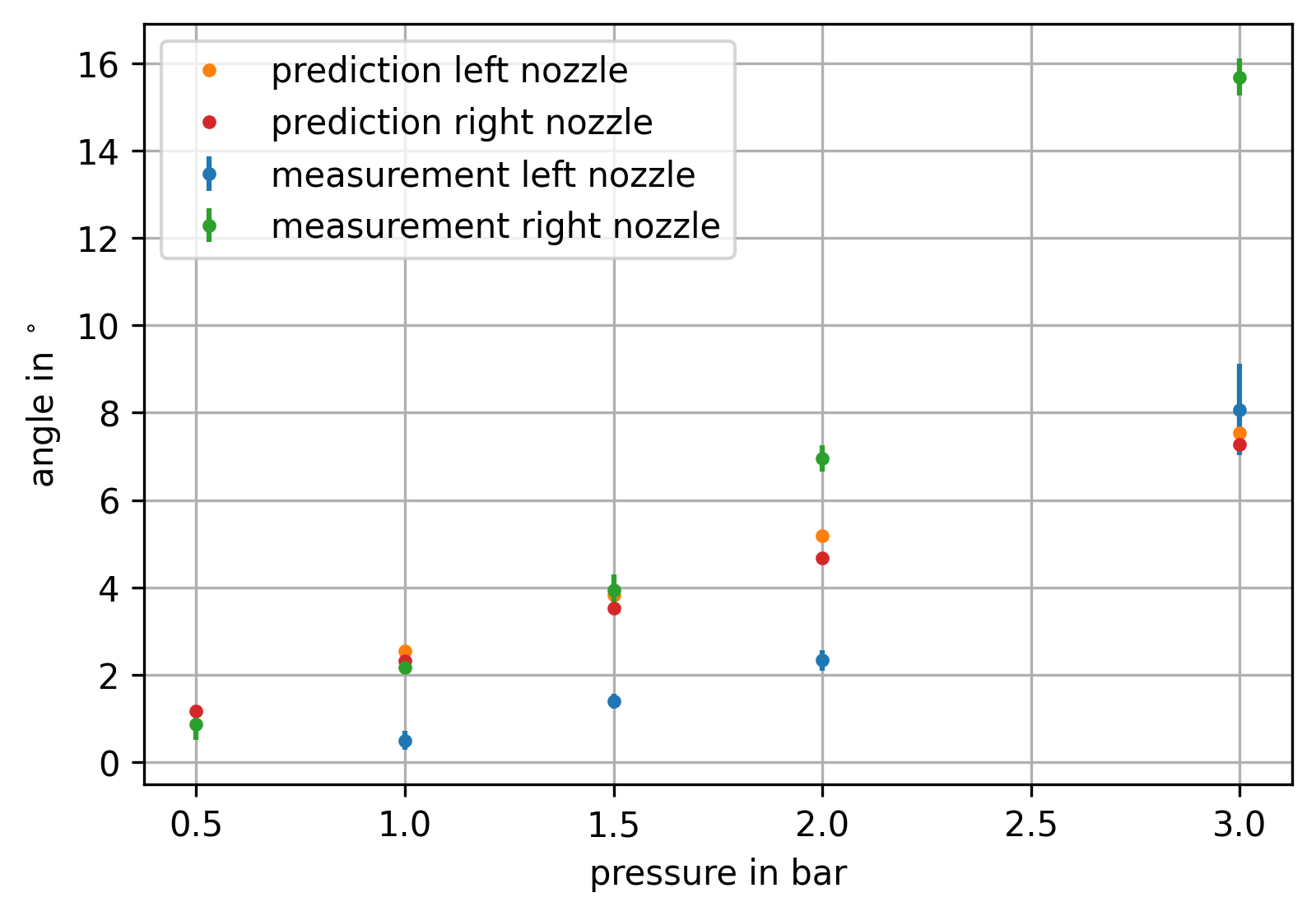}
  \caption{Measured and predicted deflection angles for different helium pressures for both deflection nozzles in figure \ref{zweiDüsen}.}
  \label{vergleich}
\end{figure}
Measurements are performed with helium pressures of $\SI{0.5}{\bar}$ to $\SI{3}{\bar}$. As two nozzles are operated in parallel, it is not possible to use such high pressures as with one nozzle in the previous chapter \ref{1}. A vacuum pressure of $p_\mathrm{ch}=\SI{2.6e-2}{\milli\bar}$ is achieved when both nozzles are operated at $\SI{3}{\bar}$. Higher helium pressures would cause the turbo pumps in the interaction chamber to reach their operating limit. \\
The bending angles are determined by aligning tangents to the beam, as shown in figure \ref{zweiDüsen} on the right. The averaged angle is plotted in dependency of the pressure for the left (blue) and right (green) nozzle in figure \ref{vergleich}. The standard deviation of the measured angles (three measurements each) is given as error bars. Additionally, the expected bending angle is calculated according to equation \eqref{alpha2} for each scenario and plotted in figure \ref{vergleich}. The predictions for both nozzles differ slightly since the nozzles do not have the same distance $d$ to the hydrogen beam. The distance $d$ is evaluated individually for each pressure setting and each nozzle.\\
The comparison between measured and expected angles shows good agreement for the right nozzle for the lower pressures. For $\SI{3}{\bar}$, a larger deviation is observed. The left nozzle seems to have a smaller deflection effect. Presumably, the gas ejected from the left nozzle does not fully hit the hydrogen beam.\\
With the simple estimation of equation \eqref{alpha2}, it is possible to predict the deflection angles of the hydrogen jet to a certain extent if the nozzle is aligned to it. It becomes also obvious if the bending gas does not hit the beam completely (left nozzle in figure \ref{vergleich} in blue). At higher bending pressures, the formula no longer seems to describe the measured angles. As the derivation is based on simplified assumptions, this is not surprising. For example, gas ejection patterns are more complex than assumed here. In reality, the gas is not ejected with the same divergence as the nozzle, and the speed and density of the gas are not constant. Especially at higher pressures, turbulent phenomena occur affecting the mentioned quantities. Furthermore, a single droplet instead of a filament is assumed in the derivation. If a cylindrical filament is considered, the effect that the string exerts a force on itself in y-direction due to its mass must be taken into account. However, as the mass is quite small, this effect does not seem to play a major role.
Despite the simplifications made here, the calculated angles match the measured ones to a certain extent. To improve the calculations, simulations of the ejection patterns of the helium nozzles are currently being developed.
At this point it has to be emphasized that the presented calculations are especially important both for the general understanding of the cryobending mechanism and for a rough estimate of the required correction nozzle pressures.\\
Based on these results, a refined system will be developed in the future.
A system consisting of four circularly arranged nozzles that can be operated at different pressures is currently being designed. By using two such arrays in series, complete control over the position of the filament in the plane perpendicular to its propagation direction can be achieved. If the ejection angle of the filament changes over time, e.g. caused by small impurities at the nozzle exit, its desired position can be maintained by adjusting the helium pressures of the nozzle array.
In a future application, e.g. in an accelerator experiment, the trajectory of the frozen filament before and after cryobending might be automatically recorded by a camera system.
Very recently, initial studies on such an automatic real-time correction system have been started and it was already possible to reconstruct the filament trajectory within less than one second. Considering the fact that a trajectory correction is needed on a timescale of minutes or longer, it is very likely that such a real-time correction system will be available soon.\\

\section{Conclusion}
A new target device for cryogenic target streams was designed, built up and used for detailed studies on the generation of stable frozen hydrogen beams in vacuum. It could be shown that hydrogen filaments with a diameter of e.g. \SI{10}{\micro\m} can be produced stably for over 100 hours, which demonstrates promising potential for the use in accelerator or high-power laser facilities. 
At a distance of \SI{2.8}{\mm} behind the target beam production nozzle, the 
position of the target beam deviates in the range of only tens of microns showing great stability. The droplet target is therefore ideal for an experiment such as MAGIX, in which the interaction point between the electron beam and the target is planned to be directly below the nozzle \cite{magixpaper}. \\
The calculated stability at a distance to the nozzle of \SI{2.1}{\m} as for the future $\mathrm{\bar{P}ANDA}$ experiment is well below $\SI{1}{\cm}$, which corresponds to the requirements of the experiment.
If the target beam does not leave the nozzle in exactly the desired direction, it may not reach the interaction point. To combat this issue, the cryobending system we propose and discuss is of great importance. Instead of moving and tilting the entire target source, correction nozzles can be implemented to manipulate the beam path by deflection with gas. As a proof of principle, we showed that the deflection of a cryogenic hydrogen beam works with one and two nozzles. With simple assumptions, it was possible to derive an analytical formula for the bending angle. The measured bending angles match with the expectations of this formula to a certain extent. Nevertheless, simulations of the gas ejection patterns will be developed in the future to increase the accuracy of the deflection angle prediction. \\
This achievement opens the door for further development. By using two arrays of circularly arranged nozzles, complete control over the target beam can be achieved in the plane perpendicular to the spread direction of the target beam. This allows the target to be swept across the accelerator beam to enable or disable the beam-target interaction instead of deflecting the accelerator beam.

\section*{Acknowledgement}
We acknowledge the excellent work of D. Bonaventura and of our mechanical and electronic workshops.
\section*{Funding}
Funding was received from GSI F\&E (MSKHOU2023), EU Horizon 2020 programme (824093), BMBF (05P21PMFP1) and NRW Netzwerke (NW21- 024-E).

\bibliography{literatur}
\bibliographystyle{elsarticle-num}

\end{document}